# What if we considered awareness for sustainable Kwowledge Management ?
## *Towards a model for self regulated knowledge management systems based on acceptance models of technologies and awareness*

Carine Edith TOURE[1,2], Christine MICHEL[1,2] and Jean-Charles MARTY[3]
[1]Université de Lyon, CNRS
[2]INSA-Lyon, LIRIS, UMR5205, F-69621, France
[3]Université de Savoie, LIRIS, UMR5205, France
{carine-edith.toure, christine.michel, jean-charles.marty}@liris.cnrs.fr

Abstract: We propose, in this paper, a model of continuous use of corporate collaborative KMS. Companies do not always have the guaranty that their KMS will be continuously used. This statement can constitute an important obstacle for knowledge management processes. Our work is based on the analysis of classical models for initial and continuous use of technologies. We also analyse the regulation concept and explain how it is valuable to support a continuous use of KMS. We observed that awareness may be a regulation means that allows taking this problem into account. Awareness is a concept, which has been profusely used to improve user experience in collaborative environments. It is an important element for regulation of activity. In our model, we assume that one can integrate awareness in information systems to positively influence beliefs about them. The final objective of our work is to refine some concepts to fit the particularities of collaborative KMS and to propose an awareness regulation process using the traces of the users' interactions with the systems.

# 1 INTRODUCTION

Using KMS to support knowledge management (KM) initiatives in companies is, nowadays, one of the most often used approaches for KM (Boughzala & Ermine, 2007). Companies increasingly invest in collaborative or cooperative information systems that promote capitalization of knowledge and interactions between actors using this knowledge through the system (Ermine, 2008). A successful corporate KM process will thus maintain continuous interactions between users and the system. The core functionalities of KMS being publication, discovery, collaboration and learning (Maier, 2007), collaborators must publish/share, seek for information, collaborate via the KMS in order to sustain the knowledge flow within the system.

Nevertheless, this is not always the case and companies usually have to deal with problems of acceptance and use of their KMS.

The acceptance of a system can occur only when the initial acceptance has been considered. Initial acceptance corresponds to the first effective use of the system. The acceptance is then satisfied when users realize continuous use of the system, this is called *continuance* (Bhattacherjee, 2001). Thus, to carry out a sustainable KM initiative, companies have to ensure a continuous use of their KMS.

In this paper, we propose to address the general issue of regulation while using corporate collaborative KMS. By regulation, we mean the sustained commitment of the users toward the KMS that guarantees the effective and long-term sharing, seeking, learning and collaboration within users of the company via the system.

It has been proven that activity awareness has a prominent role in improvement and regulation of interactions between the users and the information system (Antunes, Herskovic, Ochoa, & Pino, 2014) (Carroll et al., 2011). Being in the context of collaborative systems, we will propose a model, for self-regulation and sustainable KMS, which takes into account the concept of awareness in the use activity. Our paper is organized as follows: in section 2, we first propose to present an overview of models for initial acceptance and continuance, and then we discuss them and finally propose our model for self-regulation of systems. In the last section, we conclude and provide possible directions for future research.

## 2 BACKGROUND AND PROPOSITION

Most of acceptance models that have been published in literature derived from social psychology theories. They propose to explain people behaviours. These researches led to models that are designed according to a pattern published in the book of (Fishbein & Ajzen, 1975). This pattern, as shown in the above picture, contains causal links between the beliefs about an object, the attitudes toward it, and the intentions and behaviours associated to it. In the following, we present some of the core models of acceptance.

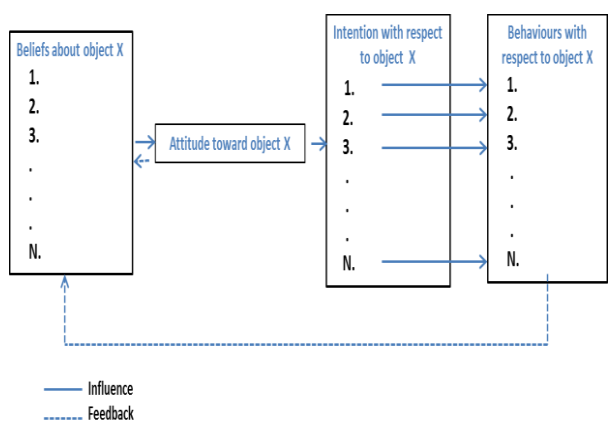

Figure 1: View of the pattern showing causal relationships between beliefs, attitude, intention and behaviours

### 2.1 Initial use of systems

In order to predict the effective use of information systems in corporate environments, (Davis, 1993) proposed the *Technology Acceptance Model* (TAM). In the TAM model, the beliefs of perceived usefulness and perceived ease of use of a system are the two fundamental criteria to build a positive attitude toward technology and to stimulate the user to start first experiments. More recently, (Venkatesh et al., 2003) published the *Unified Theory of Acceptance and Use of Technology* (UTAUT) model, a unified model of 8 reference models and theories for acceptance. About a decade later, (Venkatesh, Thong, & Xu, 2012) proposed the UTAUT2 which is an updated and more complete version of their previous model. In addition to factors like performance expectancy or effort expectancy, they add some moderators such as the age, the gender and the experience of the users.

### 2.2 Continuous use of systems

The different models cited previously were very useful to figure out the intention of use and the effective behaviours of people regarding technologies. To address continuance, (Bhattacherjee, 2001) proposed the *Expectation Confirmation Model* (ECM) that uses variables like perceived usefulness, confirmation and satisfaction. Confirmation can be defined as the extent to which the user opinion before using the system meets the perception after an effective use. This confirmation belief, which is constructed from user experience, influences variables of satisfaction, continued use intention and effective continued use.

Likewise, the *Information System Success Model* (ISSM) of (Delone, 2003) also takes the use of the system as a factor for a successful acceptance process. The ISSM model indeed considers that the system quality, the information quality, and the service quality are to be considered independently. They condition differently the intention of use and also the satisfaction and the net benefits, ensuring continuance.

(Jennex & Olfman, 2004) have adapted the ISSM model for KMS systems by considering knowledge quality and have validated it use, for KMS use evaluation.

### 2.3 Discussions

We can observe a good level of coherence between models of initial acceptance and those of continuance. Indeed, the process of acceptance begins with first beliefs that can be generated by external stimuli like system quality or information quality (cf. TAM or UTAUT models). Those beliefs impact the user's attitude toward the system, the intention of use and therefore the effective behaviour of use. After, this initial cycle of use, the user acquires an experience that helps him to construct a new belief confirming or disconfirming the previous ones. This confirmation/disconfirmation thus impacts his/her attitude (satisfaction or dissatisfaction) and intention of use in the future, and so on.

We can thus infer a spiral model (fig.2) of sustainable use of systems based on the models pattern of acceptance and continuance. The system with its functionalities (information sharing, discovering, publishing and learning) is in the centre of the figure. When the user is first confronted with the system, it is not necessarily through a use. It can be through a presentation, a talk or an advertisement. This first confrontation will influence emerging beliefs (e.g. perceived usefulness, performance expectancy, or effort), attitudes/intentions (e.g. satisfaction, use intention) and behaviours (e.g. initial use, continuance). Then s/he chooses to use it or not, to build a new experience with it, to confirm or not its beliefs, and so on as we explained before. The continued use thus depends on the results of each phase.

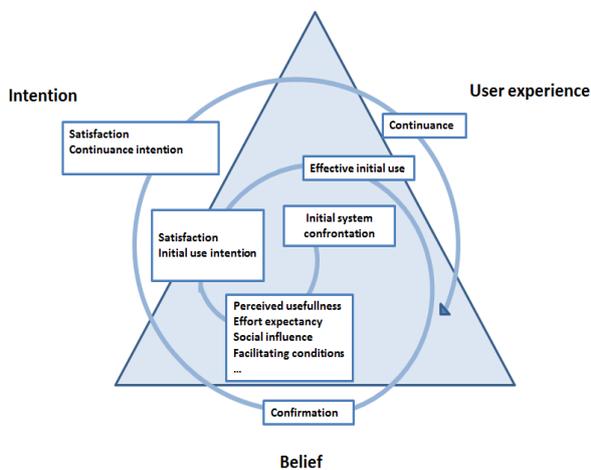

Figure 2: View of our synthetic model for sustainable use of systems

(Fishbein & Ajzen, 1975) notice in their book that attitudes and intentions are constructed at the same time in someone brain. So, in order to simplify our model, we merged the attitude and intention variables under a unique label that is intention. We also replaced the use behaviour by the user experience to express that use have an impact on user. This synthetic model will inspire us to propose a model of sustainable use of KMS integrating awareness concept. Our objective is to find a means to reinforce positive beliefs and satisfaction in order to sustain the continuance of use. We make the hypotheses that awareness functions can be helpful to do that.

## 2.4 Considering awareness to regulate the use of KMS

### 2.4.1 Awareness in collaborative systems

The concept of awareness has been used for many years in the domain of computer supported collaborative work (CSCW). (Harrison & Dourish, 1996) defines awareness as: *"The sense of other people's presence and the ongoing awareness of activity which allows us to structure our own activity, seamlessly integrating communication and collaboration ongoingly and unproblematically."* It helps to support and facilitates interactions between the users and the system (Antunes et al., 2014). It also helps the user to construct the requisite knowledge for performing his/her complex tasks (Gutwin & Greenberg, 2002). Being aware of the actual activities allows people to take autonomous decisions for problem resolution. Awareness has been profusely used to improve user experience in collaborative environments. It allows the effective regulation between actors participating in a shared activity (Carroll et al., 2011). There are various types of awareness (Antunes et al., 2014). Indeed awareness can point out some specific elements about the activity: *collaboration awareness, location awareness, context awareness, social awareness, workspace awareness, situation awareness, metacognitive awareness.* Awareness can also reflect the activities of a particular person or a group of people: *group awareness, individual awareness.*

Awareness is used to support reflexive practices. In his study of professional practises, (Schön, 1987) showed that reflexive thought is a continuous cognitive process, in which knowledge appears through an iterative thinking process. A reflexive process allows learners to be conscious of what they have to do and how they do it, to analyse their learning processes, to change and adapt their behaviours in order to improve their way of learning. The awareness of the action being performed then becomes the source of knowledge and learning. The group awareness has been defined by (Janssen, Erkens, & Kirschner, 2011) as knowledge about the social and collaborative environment the person is working in (e.g., knowledge about the activities, presence or participation of group members). The authors argue that group awareness tools supply information to users to facilitate coordination of activities in the content space (space of collaboration where users exchange information, discuss or solve problems) or the social space (space for positive group climate, effective and efficient collaboration).

The model of (Krogstie, Schmidt, & Mora, 2013) describes the links between reflection and knowledge in professional contexts.

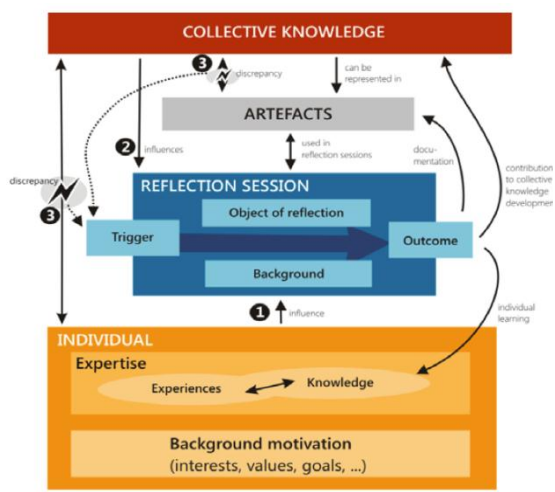

Figure 3: A model connecting knowledge and reflection (Krogstie et al., 2013)

Reflection sessions are supported by the visualisation of indicators of the activities or the state of mind. In some cases, the indicators are presented directly near the activities they reflect. NAVI surface (Charleer, Klerkx, Santos, & Duval, 2013) for example presents visualizations of user's communication activities by a "badge" presentation. In other case, indicators are presented globally, into a dashboard as it is the case in (Ji et al., 2013). The reflection can be done by the user him/herself or collaboratively, guided by an animator. In the individual and group cases, the change process occurring into this reflection session is called a regulation process.

Regulation is defined as the *"individual and social processes of adaptation, engagement, participation, learning, and development."* The authors also introduce Self-regulation as *"the cognitive and metacognitive regulatory processes used by individuals to plan, enact, and sustain their desired courses of action"*(Volet, Vauras, & Salonen, 2009). Self-regulation is defined by (Zimmerman, 2000) as a three steps process: self-monitoring, self-judgement, and self-reaction.

### 2.4.2 Awareness for knowledge management systems

In collaborative KMS perspective, awareness can be useful to encourage users to publish and share information. They can also be informed of new sharing and updates, and thus improve access to recent content. This is an improvement for discovery, publication and collaboration functions of KMS systems. For example, in a collaborative process of submission/publication of articles in a corporate blogging platform, it is really useful for both the contributor and the validator to have pieces of information about the status of their activity. The contributor will need to know whether or not his/her new articles are actually processed by the validator, who will need to have an overview of all the submissions s/he has to validate (Gendron, 2010).

In addition, *metacognitive awareness* can improve the learning process (Peña, Kayashima, Mizoguchi, & Dominguez, 2013), which is another KMS core functionality. Indeed, indicators of cognitive awareness can for example present to the learner his/her knowledge level or the improvement of knowledge, the most difficult or easiest knowledge and the number of solutions proposed by each learner (Ji et al., 2013).

We thus assume that by proposing awareness functionalities within the KMS, we could get a better support of the regulation process and improve the whole cycle of KMS use. Our model of continuous use of systems who integrates awareness function is presented in fig.4.

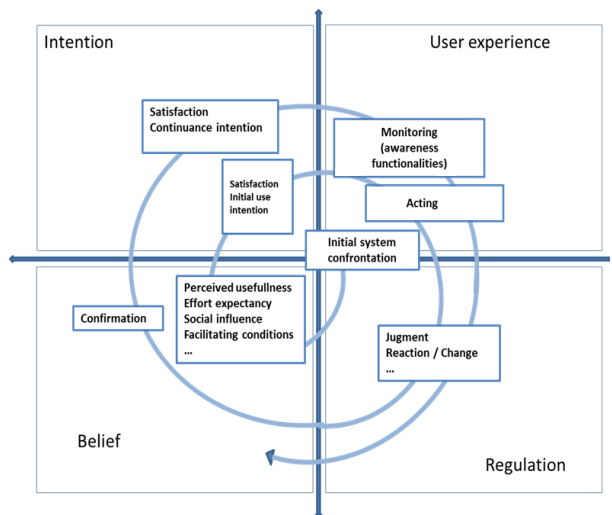

Figure 4: View of the model for sustainable use of KMS systems with awareness functions.

This cycle for sustainable use of KMS begins, as our synthetic model presented previously, with an initial confrontation of the user with the system. This will incent the emergence of different beliefs (perceived usefulness, effort expectancy, etc.), and intentions. To improve the user experience phase, we will add some awareness functionalities in the system. When the user will use the system, he will become aware of collaboration and publication done by the others or by him/herself. Moreover, awareness functions on KMS can improve the discovery and learning processes. These interactions with the system will promote judgments, reaction and changes in the behaviours of the user. The user will adapt him/herself according to the system. We assume that this phenomenon of regulation can positively influence beliefs of confirmation, then intentions (e.g. satisfaction, continued use intention) and behaviours (continued use).

According to ECM model, these beliefs evolution can reinforce the user's engagement and participation. Awareness indicators also help the user to improve his/her skills, because s/he learns from a global view on available information. His/her behaviour can thus change. As an example, the (Jennex & Olfman, 2004) KMS success model presents the perceived benefit belief that positively influences the continued KMS use.

# 4 CONCLUSION AND FUTURE WORK

In this work, we investigated awareness as an incentive in the process of continuous use of systems. We first reviewed several models of acceptance of technologies and information systems. This preliminary work helped us to deduct a synthetic model of acceptance that inspired us to propose a model for sustainable use of collaborative KMS. We integrated in this model the concept of awareness to emphasize users' beliefs and reinforce continuous use of the system.

Nevertheless, our work has just begun and we can identify a number of steps we still have to achieve. First, this model is based on researches made for information systems and CSCW. As we are interested in continuous use of collaborative KMS, we need to precise, according to characteristics of KMS, what sub-elements of each square are relevant for our context. This will lead us to analyse more deeply, each variable of the different identified models, and keep only those that are valuable. Then, we will implement and then evaluate our model. Indeed, we are working with the Société du Canal de Provence (SCP), a hydraulics Services Company located in the Provence Alpes Côte d'Azur French region. This company has massively invested in a KMS for about a couple of decade. But the initiative didn't prove a great success because of a lack of use. We want to add several awareness functionalities in their KMS, and thanks to activity indicators calculated with activity traces (Karray, Chebel-Morello, & Zerhouni, 2014), we will hopefully observe and measure the usage of the system.

Finally, we also aim, based on this model, at designing a cyclic methodology for implementation of collaborative KMS that are self-regulated thanks to awareness functionalities within the system.